\documentclass[useAMS,usenatbib]{mn2e} 
\usepackage{graphicx,epsfig} 
\usepackage{epsf}
\usepackage{aas_macros}
\usepackage{amsmath}
\usepackage{amssymb}
\usepackage[usenames]{color}

\newcommand{\kepler}{\textit{Kepler}}

\title[Super-Nyquist Asteroseismology]{Super-Nyquist asteroseismology with the Kepler Space Telescope}

\author[S.J. Murphy et al.] 
{Simon J. Murphy$^{1}$\thanks{email: smurphy6@uclan.ac.uk}, Hiromoto Shibahashi$^{2}$, Donald W. Kurtz$^{1}$
\\
\\
$^{1}$Jeremiah Horrocks Institute, University of Central Lancashire, Preston PR1 2HE\\
$^{2}$Department of Astronomy, The University of Tokyo, Tokyo 113-0033, Japan\\
}

\begin{document}

\maketitle 

\begin{abstract}
Barycentric corrections made to the timing of \kepler\ observations, necessitated by variations in light arrival time at the satellite, break the regular time-sampling of the data -- the time stamps are periodically modulated. A consequence is that Nyquist aliases are split into multiplets that can be identified by their shape. Real pulsation frequencies are distinguishable from these aliases and their frequencies are completely recoverable, even in the super-Nyquist regime, that is, when the sampling interval is longer than half the pulsation period. We provide an analytical derivation of the phenomenon, alongside demonstrations with simulated and real \kepler\ data for $\delta$\,Sct, roAp, and sdBV stars. For {\it Kepler} data sets spanning more than one {\it Kepler} orbital period (372.5\,d), there are no Nyquist ambiguities on the determination of pulsation frequencies, which are the fundamental data of asteroseismology.
\end{abstract}

\begin{keywords}
asteroseismology -- methods: data analysis -- stars: oscillations -- stars: variables: $\delta$\,Scuti -- stars: variables: general
\end{keywords}

\section{Introduction}

It is well known that all light emitted or received is Doppler shifted by the motion of the emitting or receiving body. In astronomy the Doppler shift from the emitting body is used to deduce radial velocity; the Doppler shift of the receiving body, the Earth, is removed from the observations by correcting the times of observation to the Solar System barycentre. All of this is standard practice. The Doppler shift from emitting bodies has traditionally been used to deduce orbital motion for binary stars and for exoplanets with spectroscopic radial velocities, and with $O-C$ analyses in the cases of binary pulsars and orbiting pulsating stars. Recently, \citet{shibahashi&kurtz2012} have shown how this orbital Doppler shift can be seen directly in the Fourier transform of light variations in binary pulsating stars and used to derive the mass function without the need of spectroscopic observations. Now, we show here, both analytically and in practice, that the Doppler shift at the receiving body -- in this case the {\it Kepler} Space Telescope -- allows the unique identification of periodic frequencies in the emitting body without any Nyquist ambiguity.

The Nyquist frequency of equally-spaced data represents an upper limit on a range of frequencies over which a Fourier transform is unique. It is well-defined as $f_{\rm Ny} = 1/2\Delta t$, where $\Delta t$ is the sampling interval between two consecutive points; this is also known as the ``cadence'' of the data. In perfectly equally-spaced data there is an infinity of frequencies that fit the data equally well. External physical constraints are mandatory to select the frequency range that is appropriate for pulsating stars. Commonly, the cadence is selected so that the known plausible frequency range for the type of star lies in the range $[0, f_{\rm Ny}]$, and the infinity of ambiguous higher frequency Nyquist aliases are ignored as unphysical. 
 
However, it is not possible to adjust the observing cadence for {\it Kepler} mission data to accommodate the study of pulsating stars in different frequency regimes. The {\it Kepler} mission's prime goal of detecting Earth-like planets in the habitable zone led to a design where \kepler\ data are available in two cadences: long-cadence (LC) at 29.43\,min, and short-cadence (SC) at 58.85\,s, with corresponding Nyquist frequencies of 24.469\,d$^{-1}$ and 734.07\,d$^{-1}$, respectively. The number of short-cadence observing slots is limited to 512 by telemetry constraints, and most of those observing slots are needed for higher time resolution studies of exoplanet transits. With this limitation most classical pulsators observed by {\it Kepler} have used LC, even though many such pulsators --  such as $\delta$\,Sct, roAp, sdBV, pulsating white dwarf and $\beta$\,Cep stars -- have pulsation modes with frequencies that can exceed the LC Nyquist frequency. 
\citet{murphy2012} showed some of the difficulties encountered with Nyquist aliases when studying $\delta$\,Sct stars with \kepler\ LC data.

Now, we show in this paper how the correction of the times of observations of {\it Kepler} data to the Solar System barycentre generates a time-dependent Nyquist frequency, and that this completely removes all Nyquist alias ambiguities in the amplitude spectrum. For data sets longer than one {\it Kepler} orbit, true pulsation frequencies can always be distinguished from all Nyquist aliases. The observing cadence for {\it Kepler} data is not a barrier to asteroseismic modelling of pulsation frequencies because of Nyquist ambiguity -- there is none. In practice, of course, higher frequencies have reduced amplitude with LC data (see \citealt{murphy2012}), and that affects the signal-to-noise ratio, which decreases with increasing frequency. 

The \kepler\ spacecraft is in a 372.5-d, heliocentric Earth-trailing orbit. Four times per orbit the satellite must perform a roll to keep its solar panels facing the Sun, so the data are divided into Quarters (denoted Qn) and a brief gap in observations occurs. In addition to the data downlink that takes place at each quarterly roll, two more downlinks occur at approximately 30-d intervals in the middle of each quarter, effectively dividing the data into three `months' per quarter (denoted Qn.m for SC data\footnote{LC data are not separated into different files around these months like the SC data are, but the gaps are still present.}). These events represent the most frequent gaps in \kepler\ data which are otherwise nearly continuous at their micromagnitude precision. Other gaps due to occasional safe-mode events have occurred, and we refer the reader to the Data Characteristics Handbook\footnote{available at http://keplergo.arc.nasa.gov/Documentation.shtml} for more information. However, the important property of the observations with respect to the gaps is that observations are always taken at regular intervals according to the spacecraft's clock, which can be expressed as some integer product with $\Delta t$ from some (fixed) arbitrary start time, i.e. $t_n = t_0 + n\Delta t$. In this regard, the \kepler\ data may be described as equally spaced.

Timing onboard the spacecraft is in Julian Date, but the spacecraft is in orbit around the solar-system's barycentre and as such the arrival times of photons from the \kepler\ field are shifted seasonally by up to $\pm 200$\,s from those of the Solar System barycentre. Although data are sampled regularly onboard, the time stamps are subsequently modified when converted into Barycentric Julian Date (BJD) and hence become irregularly sampled: no longer can the observation times be represented in the form $t_n = t_0 + n\Delta t$ for integer values of $n$. The effect on super-Nyquist frequencies has been observed by \citet{baranetal2012}, and described as a `smearing' of the frequencies. In this paper we show how the satellite motion results in multiplets being generated out of the Nyquist aliases, and how this can be used to distinguish these aliases from real pulsation frequencies.

\section{Analytical Derivation}
\label{sec:2}
\subsection{Discrete Fourier transform}
\label{sec:2.1}
\subsubsection{General description}
\label{sec:2.1.1}
Let us consider first a general case of discrete data $\{x(t_n)\}$, which can be regarded as the following function of any value of $t$;
\begin{equation}
	x_{N}(t) \equiv \sum_{n=0}^N x(t)\delta(t-t_n) ,
\label{eq:2.1}
\end{equation}
where $x(t)$ on the right-hand side is a continuous function $x(t)$ and $\delta(t-t_n)$ is Dirac's delta function. It is well known that the Fourier transform of $x_N(t)$ is expressed as a convolution of the Fourier transform of $x(t)$ and the sampling window spectrum $W_N(\omega)$: 
\begin{eqnarray}
	{{1}\over{N+1}} F_N(\omega)
&\equiv&  
	{{1}\over{N+1}} \int_{t_0}^{t_N} x_{N}(t)\exp(i\omega t)\,dt
\nonumber \\
&=&
	(F * W_N )(\omega),
\label{eq:2.2}
\end{eqnarray}
where
\begin{equation}
	F(\omega) \equiv \int_{-\infty}^\infty x(t)\exp(i\omega t)\,dt
\label{eq:2.3}
\end{equation}
is the Fourier transform of the continuous function $x(t)$
and
\begin{equation}
	W_N(\omega) \equiv {{1}\over{N+1}} \sum_{n=0}^N \exp(i\omega t_n) .
\label{eq:2.4}
\end{equation}
Here, we have used the following expression of the delta function:
\begin{equation}
	\delta(t)=\int_{-\infty}^{\infty} \exp(-i\omega t)\,d\omega.
\label{eq:2.5}
\end{equation}
That is, the Fourier transform of discretely sampled data is equal to the convolution of the Fourier transform of the original continuous function and the window function. If we assume $x(t) = \cos(\omega_0 t + \phi)$, i.e. pulsation with a single mode of angular frequency $\omega_0$, regarding this as a model of a pulsating star, then  
\begin{equation}
	F(\omega)={{1}\over{2}} \left\{ \delta(\omega+\omega_0) + \delta(\omega-\omega_0) \right\} .
\label{eq:2.6}
\end{equation}
Thus the Fourier transform of the sample series $F_N(\omega)$ consists of a superposition of the shape of the window spectrum, shifted by $\pm \omega_0$. 

\subsubsection{The case of a uniform cadence}
\label{sec:2.1.2}
\begin{figure}
\begin{center}
\includegraphics[width=\linewidth]{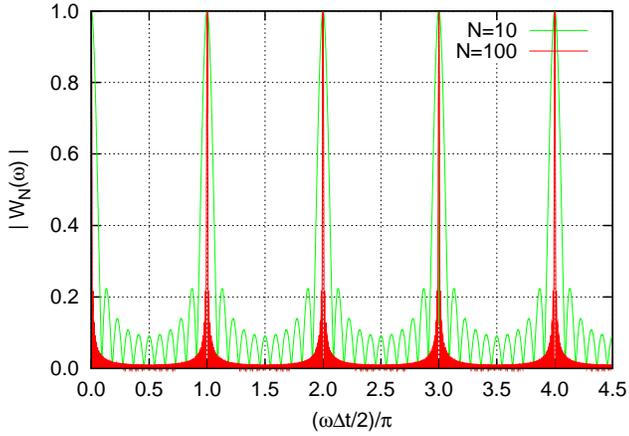}
\end{center}
\caption{The sampling window spectrum $W_N(\omega)$ in the case of sampling with a constant time interval. The abscissa is $\omega\Delta t/(2\pi)$, where $\Delta t$ is the sampling time interval. The red curve shows the case of $N=100$ data points, while the green curve shows the case of $N=10$. The window function shows conspicuous peaks at $\omega=n \omega_{\rm S}$, where $\omega_{\rm S}=2\pi/\Delta t$ is the sampling angular frequency and $n$ denotes integers.}
\label{fig:2.1}
\end{figure}

If the sampling is taken with a uniform cadence with a time interval $\Delta t$, then $t_n = t_0 + n\Delta t$, and 
\begin{equation}
	|\,W_N(\omega)\,| 
=
	{{1}\over{N+1}}
	\left| {{\sin\{(N+1)\omega\Delta t/2\}}\over{\sin(\omega\Delta t/2)}} \right| .
\label{eq:2.7}
\end{equation}
The window spectrum $|W_N(\omega)|$ has 
a series of sharp, high peaks at $\omega = n\,\omega_{\rm S}$,
where
\begin{equation}
	\omega_{\rm S} \equiv {{2\pi}\over{\Delta t}}
\label{eq:2.7a}
\end{equation}
and $n$ denotes integers (see Fig.\,\ref{fig:2.1}). 
Then $F_N(\omega)$ has  
sharp peaks at $\omega = n\, \omega_{\rm S} \pm \omega_0$. These apparent multiple peaks are the Nyquist aliases, and they look identical except for their frequencies. 

The peak in the range of $[0, \omega_{\rm S}/2]$ corresponds to the true angular frequency as long as $2\omega_0 < \omega_{\rm S}$ (here, we take $\omega_0>0$). Hence, for a given angular frequency $\omega_0$, if we select the sampling rate $\Delta t < (2\pi/\omega_0)/2$, we can identify uniquely the true frequency from the Fourier transform of the sampled data. This condition of $\Delta t < (2\pi/\omega_0)/2$ is a sufficient condition for unique identification of the true frequency, not a necessary condition.

Now let us consider a case that the sampling rate is given in advance. In this case, if the true angular frequency is lower than $\omega_{\rm S}/2$, the lowest frequency peak of the Fourier transform of the sampled data gives the true angular frequency. Note that this is again a sufficient condition, not a necessary condition. The Nyquist frequency is this frequency limit, under which we can identify uniquely the true angular frequency:
\begin{equation}
	f_{\rm Ny} \equiv {{1}\over{2}} \, {{\omega_{\rm S} }\over{2\pi}} = {{1}\over{2\Delta t}} .
\label{eq:2.8}
\end{equation}

The reason why the true frequency is not uniquely identified without external constraints in the case of sampling with a uniform cadence is the fact that all the Nyquist aliases look identical and are indistinguishable. Inversely, if we could distinguish the individual Nyquist aliases, we would be able to identify uniquely the true frequency even if $\omega_0 > \omega_{\rm S}/2$. We will see in the case of {\it Kepler} data this is indeed possible. 

\subsubsection{Sampling with a periodically modulated time interval}
\label{sec:2.1.3}

In the case of {\it Kepler}, observations are taken at regular intervals according to the clock on-board the spacecraft. Data sampling is made at 
\begin{equation}
	t_n = t_0 + n\Delta t ,
\label{eq:2.9}
\end{equation}
where $n= 0, ..., N$, and $\Delta t$ is a constant. The long-cadence (LC) data have $\Delta t = 29.4\,{\rm min}$. But, since {\it Kepler} is orbiting around the Solar System barycentre, the resulting annual variation in the distance between the stars and the spacecraft leads to modulation in the phase of observed stellar pulsation, and the time stamps are converted to barycentre arrival times. As a consequence, the time stamp interval of the {\it Kepler} data is periodically modulated.
\begin{figure}
\begin{center}
\includegraphics[width=\linewidth]{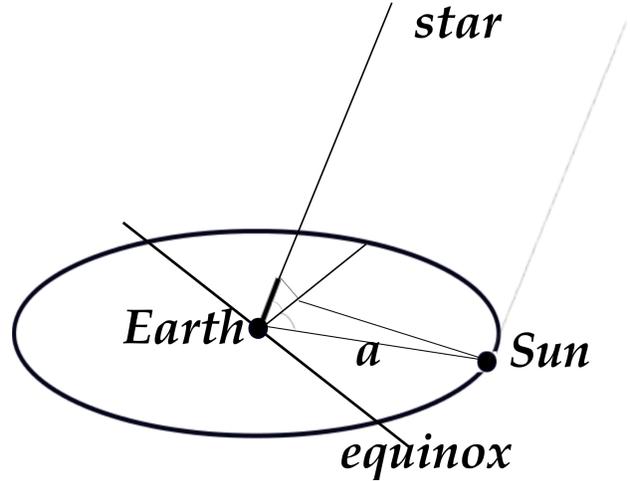}
\end{center}
\caption{A schematic picture showing geometric relations among the Earth, the Sun, and the star. The Sun is moving around the Earth on the ecliptic plane. The geocentric ecliptic longitude is measured from the vernal equinox. The star's position is $(\lambda, \beta)$, and the geocentric ecliptic longitude of the Sun is $\lambda_\odot(t)$. The difference in the path length is shown with a thick segment. The path difference is obviously given as $a\cos(\lambda - \lambda_\odot)\cos\beta$.}
\label{fig:2.2}
\end{figure}

To make the modelling simple, we ignore in this section the difference between the barycentre and the heliocentre. Also, we approximate {\it Kepler}'s orbit as a circle with a radius of $a=1\,{\rm au}$. Then, the arrival time of the light from the star to the spacecraft delays from that to the heliocentre by
\begin{equation}
	\delta t (t)={{a}\over{c}}\cos\beta\cos\{\lambda_\odot (t)-\lambda\},
\label{eq:2.10}
\end{equation}
which is often called the R{\o}mer delay, in honour of Ole R{\o}mer, who discovered observationally the fact that the speed of light is finite by comparing the observed timing of the eclipses of Jupiter's satellite Io with the estimate made at his time \citep{sterken2005}. Here, $\lambda$ and $\beta$ are the ecliptic longitude and latitude of the star, $\lambda_\odot (t)$ is the geocentric ecliptic longitude of the Sun, and $c$ denotes the light speed (see Fig.\,\ref{fig:2.2}). With the present assumption, 
\begin{equation}
	\lambda_\odot(t) = \Omega (t-t_0),
\label{eq:2.11}
\end{equation}
where $\Omega\equiv 2\pi/365\,{\rm rad}\,{\rm d}^{-1}$ and $t_0$ is the vernal equinox passage time of the Sun. It is instructive to write down here, for later use, the light time with the equatorial coordinates,
\begin{equation}
	\delta t (t)
	=
	{{a}\over{c}}
	\cos\lambda_\odot\cos\alpha\cos\delta 
	+\sin\lambda_\odot
	(\cos\varepsilon \sin\alpha \cos\delta + \sin\varepsilon \sin\delta),
\label{eq:2.12}
\end{equation}
where $\alpha$ and $\delta$ are the right ascension and the declination of the star and $\varepsilon$ is the obliquity of the ecliptic.

\begin{figure}
\begin{center}
\includegraphics[width=\linewidth]{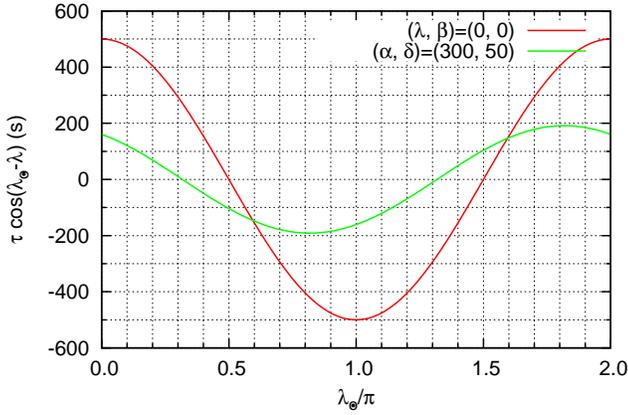}
\end{center}
\caption{The time correction for the case of $\lambda =0^\circ, \beta=0^\circ$ (red) and for the {\it Kepler} field-of-view $(\alpha=300^\circ, \delta=50^\circ)$  (green) as a function of the Sun's ecliptic longitude $\lambda_\odot$.}
\label{fig:2.3}
\end{figure}

By taking account of this light time effect, the {\it Kepler} data are stored as a function of the arrival time at the heliocentre\footnote{Strictly speaking, the data are stored as a function of the Barycentric Dynamical Time (TDB) referenced to the Solar System barycentre. The difference between the time referenced to the heliocentre and that to the Solar System barycentre is $\lesssim 4\,{\rm s}$. The main cause of this difference is the acceleration of the Sun due primarily to Jupiter and Saturn.}:
\begin{equation}
	t_{\odot, n} \equiv t_n + \tau \cos(\Omega t_n - \lambda),
\label{eq:2.14}
\end{equation}
where
\begin{equation}
	\tau \equiv {{a}\over{c}}\cos\beta 
\label{eq:2.15}
\end{equation}
gives the amplitude of modulation of the time stamps of the {\it Kepler} data. The second term 
on the right-hand side of equation (\ref{eq:2.14}) is the heliocentric time correction. Fig.\,\ref{fig:2.3} shows this time correction (the R{\o}mer delay) for the field of $\lambda =0^\circ, \beta=0^\circ$ (red) and for the {\it Kepler} field-of-view (green). The case of $\beta=0^\circ$ gives the maximum correction; $\tau = 500\,{\rm s}$. In the case of the {\it Kepler} field, $\tau \simeq 190\,{\rm s}$.

\subsection{The window spectrum of periodically modulated sampling}
\label{sec:2.2}
Our aim is to carry out Fourier analysis of the {\it Kepler} data of pulsating stars. Since the Fourier transform of the sample series $F_N(\omega)$ is a superposition of the window spectrum, the problem is essentially what the window spectrum 
$W_N(\omega)=(N+1)^{-1}\displaystyle{ \sum_{n=0}^N\exp(i\omega t_{\odot, n}) }$
looks like in the case of periodically-modulated sampling. The problem is then how to treat the terms $\exp(i\omega\tau\cos\Omega t)$ and $\exp(i\omega\tau\sin\Omega t)$. 

\subsubsection{Bessel coefficients}
\label{sec:2.2.1}
These terms can be expressed with a series expansion in terms of Bessel functions of the first kind with integer order. With the help of Jacobi-Anger expansions
\begin{eqnarray}
	\cos(\pm \xi\cos\varphi) = J_0(\xi)+2\sum_{n=1}^\infty (-1)^n J_{2n}(\xi) \cos 2n\varphi
\label{eq:2.16}
\end{eqnarray}
and
\begin{eqnarray}
	\sin(\pm \xi\cos\varphi) = \pm 2\sum_{n=0}^\infty (-1)^n J_{2n+1}(\xi) \cos(2n+1)\varphi ,
\label{eq:2.17}
\end{eqnarray}
we easily obtain the following relations:
\begin{eqnarray}
	\cos(\vartheta \pm \xi\cos\varphi) 
&=&
	\sum_{n=-\infty}^\infty (-1)^n 
	\Big\{
	J_{2n}(\xi) \cos(\vartheta+2n\varphi)
\nonumber\\
& & 
	\mp
	 J_{2n+1}(\xi) \sin\{\vartheta+(2n+1)\varphi\} \Big\}
\label{eq:2.18}
\end{eqnarray}
and
\begin{eqnarray}
	\sin(\vartheta\pm \xi\cos\varphi)
&=&
	\sum_{n=-\infty}^\infty (-1)^m 
	\Big\{
	J_{2n}(\xi) \sin(\vartheta + 2n\varphi)
\nonumber\\
& & 
	\mp  J_{2n+1}(\xi) \cos\{\vartheta + (2n+1)\varphi \}
	\Big\} .
\label{eq:2.19}
\end{eqnarray}
Here $J_n(\xi)$ denotes the Bessel function of the first kind of integer order $n$.

\subsubsection{Analytic expression of the window spectrum}
\label{sec:2.2.2}
By applying these relations, after somewhat lengthy but straightforward manipulations, 
we eventually get 
\begin{eqnarray}
& &\hspace{-1cm} (N+1)^2	|W_N(\omega)|^2
\nonumber\\
&=&
	\sum_{k=-\infty}^\infty \Bigg|
	J_k(\omega\tau) 
	{{ \sin\{(N+1)(\omega+ k\Omega)\Delta t/2\} }\over{ \sin\{(\omega + k\Omega)\Delta t/2\} }}
	\Bigg|^2
\nonumber\\
& &
	+
	\Bigg[
	\sum_{k=-\infty}^\infty {\sum_{k'=-\infty}^\infty}' 
	J_{2k}(\omega\tau) J_{2k'}(\omega\tau)
\nonumber\\
& &
	\times\cos\,  \left\{  \,2(k-k')\Omega{{N\Delta t}\over{2}} -2(k-k')\lambda \, \right\} 
\nonumber\\
& &
	+
	\sum_{k=-\infty}^\infty {\sum_{k'=-\infty}^\infty}' 
	J_{2k+1}(\omega\tau) J_{2k'+1}(\omega\tau)
\nonumber\\
& &
	\times\cos\,\left\{ \,2(k-k')\Omega{{N\Delta t}\over{2}} -2(k-k')\lambda \,\right\}
\nonumber\\
& &
	+
	\sum_{k=-\infty}^\infty {\sum_{k'=-\infty}^\infty}
	J_{2k}(\omega\tau) J_{2k'+1}(\omega\tau)
\nonumber\\
& &
	\times
	\sin
	\Big[
	\left\{ 2\omega+\,[\,2(k+k')+1\,]\,\Omega \right\} {{N\Delta t}\over{2}} 
\nonumber\\
& &
	-\,\{2(k+k')+1\}\,\lambda
	\Big]
	\Bigg]
\nonumber\\
& &
	\times
	\Bigg|
	{{ \sin\{(N+1)(\omega+ k\Omega)\Delta t/2\} }\over{ \sin\{(\omega + k\Omega)\Delta t/2\} }}
	\Bigg|^2 ,
\label{eq:2.20}
\end{eqnarray}
where $\sum '$ means summation for $k'$ except for when $k' = k$.

\begin{figure}
\begin{center}
\includegraphics[width=\linewidth]{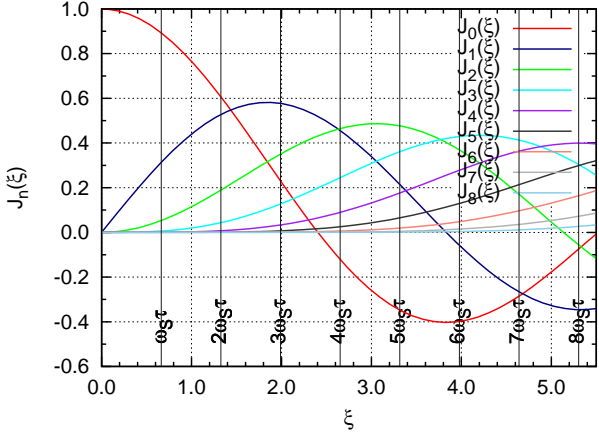}  
\end{center}
\caption{The Bessel coefficients $J_n(\xi)$ with $n=0, ..., 8$ for $\xi=[0:5.5]$.
The vertical lines show $\xi=n\,\omega_{\rm S}\tau$ $(n=0, ..., 8)$, at which the window spectrum has sharp multiplet peaks (see text).
}
\label{fig:2.4}
\end{figure}

The Bessel coefficients of different orders are orthogonal to each other (see Fig.\,\ref{fig:2.4}). Hence, among the terms on the right-hand side, the first term is dominant. Then,
\begin{eqnarray}
	|\,W_N(\omega)\,|
	&\simeq&
	{{1}\over{N+1}}
\nonumber\\
	& &\hspace{-1cm} \times
	\sum_{k=-\infty}^\infty \Bigg|
	J_k(\omega\tau) 
	{{ \sin\{(N+1)(\omega+ k\Omega)\Delta t/2\} }\over{ \sin\{(\omega + k\Omega)\Delta t/2\} }}
	\Bigg| .
\label{eq:2.21}
\end{eqnarray}
This means that the window spectrum $|W_N(\omega)|$ consists of a singlet sharp peak at $\omega=0$ and multiplets of sharp peaks at $\omega= n\omega_{\rm S} + k\Omega$, where $n$ and $k$ are integers and $\omega_{\rm S}\equiv 2\pi/\Delta t$. The amplitude of each peak is $|J_k(n\omega_{\rm S}\,\tau)|$, since $\Omega \ll \omega_{\rm S}$ (see Table\,\ref{tab:2.1}).

\begin{table}
\begin{center}
\caption{The window spectrum.} 
\label{tab:2.1}
\begin{tabular}{cclc}
\hline\hline
$n$ & $k$ & $\omega$ & Amplitude\\
\hline
$0$ & $0$ & $0$ & $1$ \\
\hline
$1$ & $0$ & $\omega_{\rm S}$ & $J_0(\omega_{\rm S}\tau)$ \\
 & $\pm 1$ & $\omega_{\rm S} \pm \Omega $ & $J_1(\omega_{\rm S}\tau)$ \\
 & $\pm 2$ & $\omega_{\rm S} \pm 2\Omega $ & $J_2(\omega_{\rm S}\tau)$ \\
 & $\cdots$ & $\cdots$ & $\cdots$ \\
\hline
$2$ & $0$ & $2\omega_{\rm S}$ & $J_0(2\omega_{\rm S}\tau)$ \\
 & $\pm 1$ & $2\omega_{\rm S} \pm \Omega $ & $J_1(2\omega_{\rm S}\tau)$ \\
 & $\pm 2$ & $2\omega_{\rm S} \pm 2\Omega $ & $J_2(2\omega_{\rm S}\tau)$ \\
 & $\pm 3$ & $2\omega_{\rm S} \pm 3\Omega $ & $J_3(2\omega_{\rm S}\tau)$ \\
 & $\cdots$ & $\cdots$ & $\cdots$ \\
\hline
$3$ & $0$ & $3\omega_{\rm S}$ & $J_0(3\omega_{\rm S}\tau)$ \\
 & $\pm 1$ & $3\omega_{\rm S} \pm \Omega $ & $J_1(3\omega_{\rm S}\tau)$ \\
 & $\pm 2$ & $3\omega_{\rm S} \pm 2\Omega $ & $J_2(3\omega_{\rm S}\tau)$ \\
 & $\pm 3$ & $3\omega_{\rm S} \pm 3\Omega $ & $J_3(3\omega_{\rm S}\tau)$ \\
 & $\pm 4$ & $3\omega_{\rm S} \pm 4\Omega $ & $J_4(3\omega_{\rm S}\tau)$ \\
 & $\cdots$ & $\cdots$ & $\cdots$ \\
\hline
$\cdots$ & $\cdots$ & $\cdots$\\
\hline
\end{tabular}
\end{center}
\end{table}

It should be stressed here that $|W_N(\omega)|$ is highly dependent on frequency $\omega$ through $\xi \equiv \omega\tau$, the argument of the Bessel function. For $\xi\ll 1$, the dominant term is only $J_0(\xi)\sim 1-\xi^2/4$ and all the higher-order Bessel functions are negligibly small. However, with the increase of $\xi$, the first-order Bessel function $J_1(\xi)$ comes larger as $J_1(\xi)\sim \xi/2$, and eventually it becomes larger than $J_0(\xi)$ around $\xi \simeq 1.5$. But the higher-order terms $J_k(\xi)$ for $k\gtrsim 4$ are still negligibly small. This means that $|W_N(\omega)|$ looks like a triplet with an equal spacing of $\Omega$ if $\omega \lesssim (2\tau)^{-1}$. With the further increase of $\xi$, the second-order Bessel function $J_2(\xi)$ also becomes non-negligible. As a consequence, $|W_N(\omega)|$ comes to look like a quintuplet. This tendency continues further: the high-order Bessel functions become non-negligible with the increase of $\omega$. Therefore, the individual Nyquist aliases are distinguishable from each other. This characteristic is different from the case of sampling with a uniform cadence. 

\subsubsection{Nyquist frequency is no longer the detection limit}
\label{sec:2.2.3}
We have shown in Sec.\,\ref{sec:2.1.1} that the Fourier transform of the discretely sampled data $x_N(t)$ is given as the convolution of the Fourier transform of the original continuous data $x(t)$ with that of the window spectrum $W_N(\omega)$. Hence the Fourier transform of discrete data with a periodically-modulated sampling interval essentially consists of multiplets around the Nyquist aliases of $W_N(\omega)$ at $n\,\omega_{\rm S} \pm \omega_0$, where $n$ denotes integers. It should be noted here that the shape of the multiplets is independent of the frequency of the mode $\omega_0$, as long as the amplitude and the frequency of the mode are stable. It should also be stressed that only the peak at $\omega=\omega_0$ with $n=0$ is a single peak, while the other Nyquist aliases are multiplets; $n\,\omega_{\rm S} \pm \omega_0 \pm k\Omega$, where $n\ne 0$ and $k$ denotes integers (see Table\,\ref{tab:2.2}). Hence, by finding the unique singlet, we can distinguish the true pulsation frequency, $\omega_0$, from its Nyquist aliases. Note also that this feature is true not only for the case of $\omega < \omega_{\rm S}/2$ but also for the case of $\omega > \omega_{\rm S}/2$.  That is, a pulsation frequency higher than the Nyquist frequency, which is still defined as $1/(2\pi\Delta t)$, is uniquely determined when the sampling rate is periodically modulated, irrespective of the number of multiples of the Nyquist frequencies crossed.\footnote{For truly irregularly (randomly) sampled data, the lowest frequency above which real frequencies are truly indistinguishable from aliases is equal to the reciprocal of the accuracy to which time is measured \citep{koen2010}. For \kepler\ this is around $10^{7}$\,d$^{-1}$, so we can distinguish aliases from real frequencies for all frequencies within the star that are physically meaningful.} The Nyquist frequency is no longer the upper limit of frequency determination in such a case. This has been mathematically proved here.
\begin{table}
\caption{Fourier transform of the data sampled with a periodically modulated interval.}
\begin{center}
\begin{tabular}{rlc}
\hline\hline
$n$ & Angular frequency & Amplitude\\
\hline
$-1$ & $-\omega_{\rm S} + \omega_0$ & $J_0(\omega_{\rm S}\tau)$ \\
 & $-\omega_{\rm S} + \omega_0\pm \Omega$ & $J_1(\omega_{\rm S}\tau)$ \\
 & $-\omega_{\rm S} + \omega_0\pm 2\Omega$ & $J_2(\omega_{\rm S}\tau)$ \\
 & $...$ & $...$ \\
\hline
$0$ & $\omega_0$ & $1$ \\
\hline
$1$ & $\omega_{\rm S} - \omega_0$ & $J_0(\omega_{\rm S}\tau)$ \\
 & $\omega_{\rm S} - \omega_0\pm \Omega $ & $J_1(\omega_{\rm S}\tau)$ \\
 & $\omega_{\rm S} - \omega_0\pm 2\Omega $ & $J_2(\omega_{\rm S}\tau)$ \\
 & $...$ & $...$ \\
 \hline
$1$ & $\omega_{\rm S} + \omega_0$ & $J_0(\omega_{\rm S}\tau)$ \\
 & $\omega_{\rm S} + \omega_0\pm \Omega $ & $J_1(\omega_{\rm S}\tau)$ \\
 & $\omega_{\rm S} + \omega_0\pm 2\Omega $ & $J_2(\omega_{\rm S}\tau)$ \\
 & $...$ & $...$ \\
 \hline
$2$ & $2\omega_{\rm S} - \omega_0$ & $J_0(2\omega_{\rm S}\tau)$ \\
 & $2\omega_{\rm S} - \omega_0 \pm \Omega $ & $J_1(2\omega_{\rm S}\tau)$ \\
 & $2\omega_{\rm S} - \omega_0 \pm 2\Omega $ & $J_2(2\omega_{\rm S}\tau)$ \\
 & $2\omega_{\rm S} - \omega_0 \pm 3\Omega $ & $J_3(2\omega_{\rm S}\tau)$ \\
 & $...$ & $...$ \\
\hline
$2$ & $2\omega_{\rm S} + \omega_0$ & $J_0(2\omega_{\rm S}\tau)$ \\
 & $2\omega_{\rm S} + \omega_0 \pm \Omega $ & $J_1(2\omega_{\rm S}\tau)$ \\
 & $2\omega_{\rm S} + \omega_0 \pm 2\Omega $ & $J_2(2\omega_{\rm S}\tau)$ \\
 & $2\omega_{\rm S} + \omega_0 \pm 3\Omega $ & $J_3(2\omega_{\rm S}\tau)$ \\
 & $...$ & $...$ \\
\hline
... & ... & ...\\
\hline
\end{tabular}
\end{center}
\label{tab:2.2}
\end{table}

\subsection{Window spectrum for the $\mbox{\boldmath$Kepler$}$ LC data}
\label{sec:2.3}
In the case of the {\it Kepler} field, $\tau \simeq 190\,{\rm s} \simeq 2.20\times 10^{-3}\,{\rm d}$. 
For the range of $\omega/(2\pi) = [\,0, 200\,]\,\,{\rm d}^{-1}$ $(\omega =\,[\,0, 1.26\times{10}^3\,]\,\,{\rm rad\,d}^{-1})$, $\omega\tau = \,[\,0, 2.77\,]\,{\rm rad}$. 
Fig.\,\ref{fig:2.4} shows the Bessel coefficients $J_n(\omega\tau)$ with $n=0, ..., 8$ for $\omega\tau=\,[\,0,5.5\,]\,$. From this figure, it is obvious that $J_n(\omega\tau)$ with $n\ge 8$ are negligibly small in this range.
\begin{figure*}
\begin{center}
\includegraphics[width=0.32\linewidth]{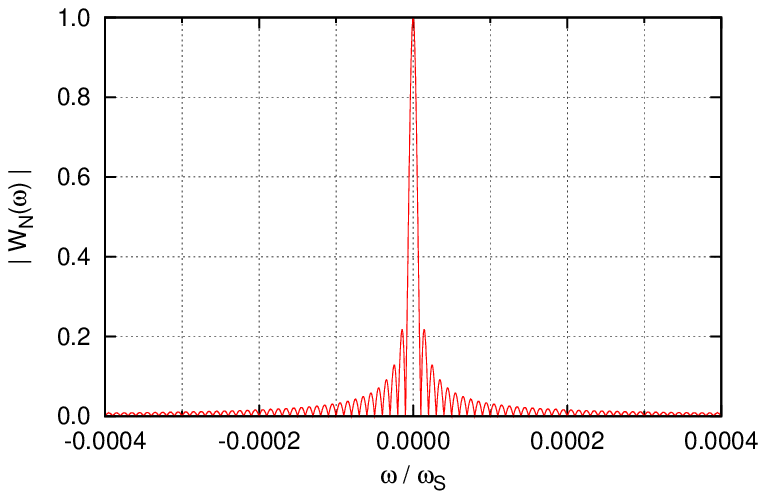}
\includegraphics[width=0.32\linewidth]{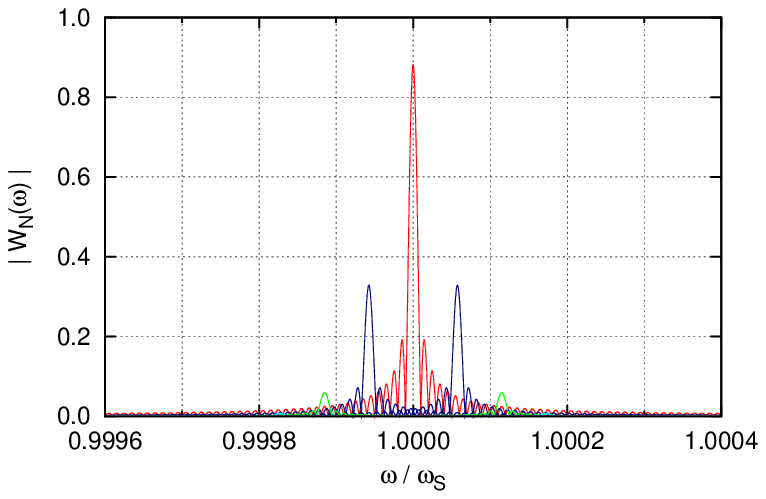}
\includegraphics[width=0.32\linewidth]{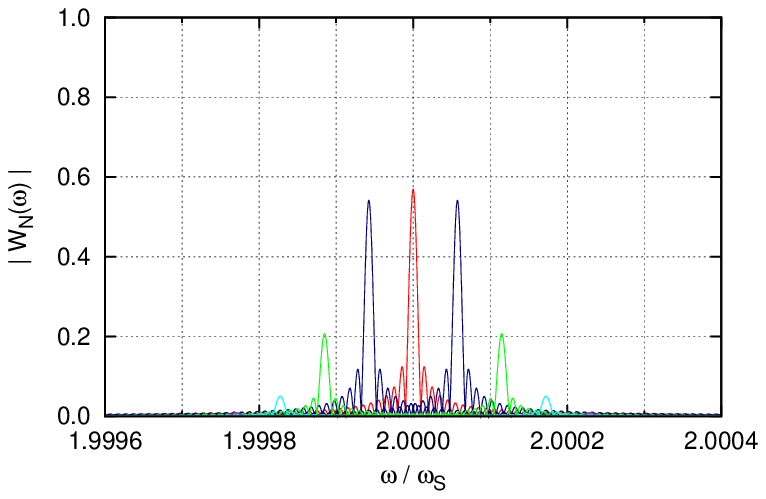}
\includegraphics[width=0.32\linewidth]{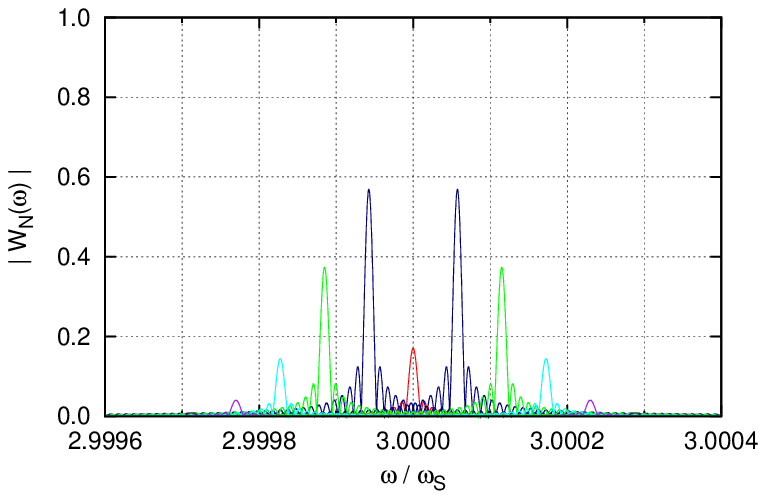}
\includegraphics[width=0.32\linewidth]{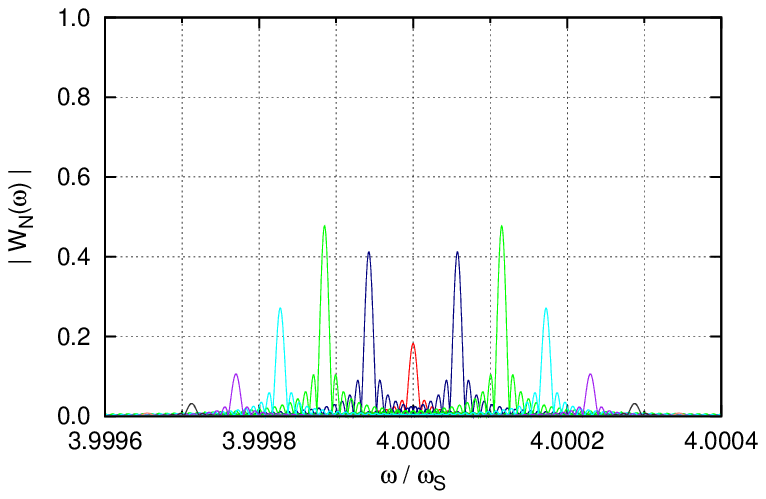}
\includegraphics[width=0.32\linewidth]{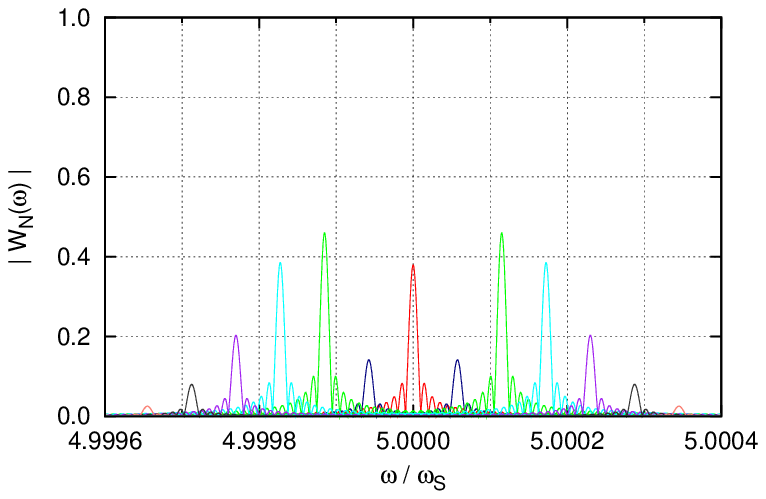}
\includegraphics[width=0.32\linewidth]{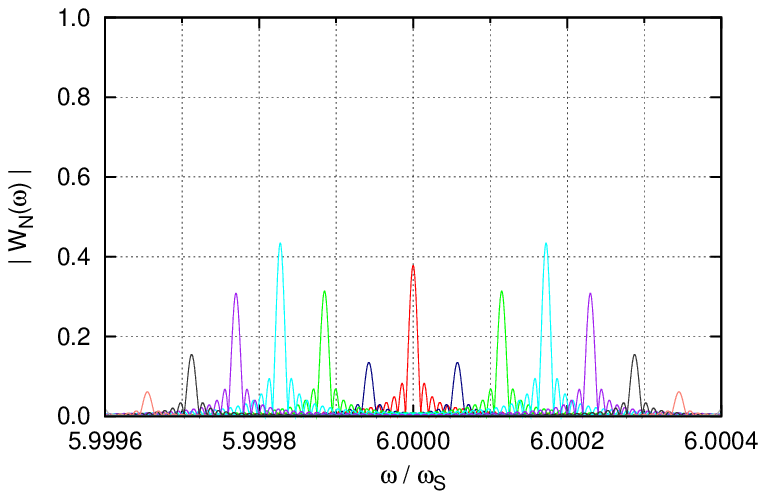}
\includegraphics[width=0.32\linewidth]{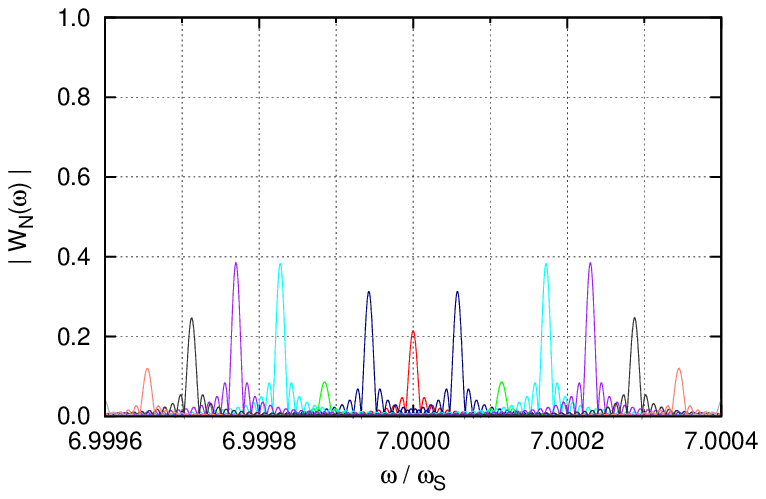}
\includegraphics[width=0.32\linewidth]{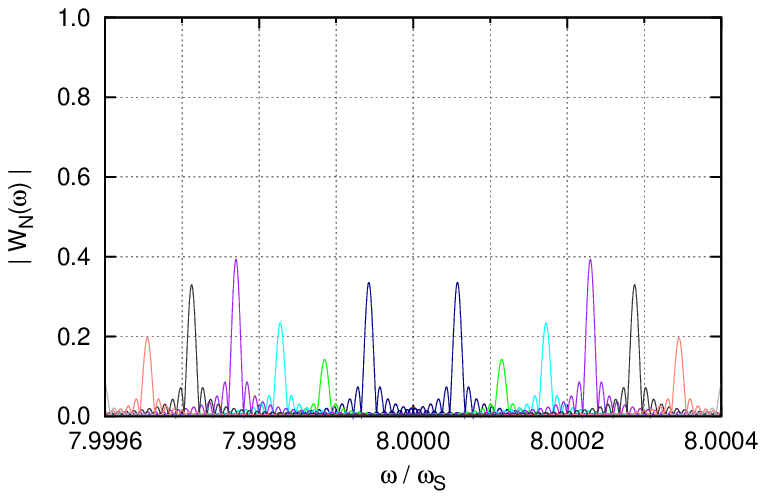}
\end{center}
\caption{The dependence of $(N+1)^{-1} |\,J_k(\omega\,\tau) \sin\{(N+1)(\omega+ k\Omega)\Delta t/2\}/\sin\{(\omega + k\Omega)\Delta t/2\}\,|$ on $\omega/\omega_{\rm S}$, for $k=0, ..., 8$, where $\omega_{\rm S}\equiv \Delta t/(2\pi)$. 
The color codes, corresponding to the order of the Bessel coefficients $k$, are in common with Fig.\,\ref{fig:2.4}.
The parameters are $\Delta t=30\,{\rm min}$; $\omega_{\rm S} = 3.016\times 10^2\, {\rm rad}\,{\rm d}^{-1}$. $\Omega=2\pi/365\,{\rm rad}\,{\rm d}^{-1}$; $\Omega/\omega_{\rm S}=5.69\times 10^{-5}$. $\tau = 190\,{\rm s} =2.20\times {10}^{-3}\,{\rm d}$; $\omega_{\rm S}\tau = 0.663\,{\rm rad}$;  $N=10^5$.
}
\label{fig:2.5}
\end{figure*}

Let us consider the case of LC of {\it Kepler}. If we take $\Delta t = 30\,{\rm min} = 2.1 \times 10^{-2}\,{\rm d}$, then $\omega_{\rm S}\equiv 2\pi/\Delta t = 3.016\times 10^2\, {\rm rad}\,{\rm d}^{-1}$. Then $\omega_{\rm S}\tau = 0.663\,{\rm rad}$. As for the modulation, $\Omega=2\pi/365\,{\rm rad}\,{\rm d}^{-1}=1.72\times 10^{-2}\,{\rm rad}\,{\rm d}^{-1}$, then $\Omega/\omega_{\rm S}=5.69\times 10^{-5}$. Parameters for the \kepler\ LC sampling are given in Table\,\ref{tab:2.3}.

Fig.\,\ref{fig:2.5} shows the dependence of $|\,J_k(\omega\,\tau) \sin\{(N+1)(\omega+ k\Omega)\Delta t/2\}/\sin\{(\omega + k\Omega)\Delta t/2\}\,|$ on $\omega/\omega_{\rm S}$ for $k=0, ..., 8$. Here, we assumed $N=10^5$, that is, the observational time span was assumed to be $\sim 5.75\,{\rm yr}$. In such a case, the high peaks are so sharp that their self-cross terms dominate over the other cross terms. As seen in Fig.\,\ref{fig:2.5}, the high peaks appear as multiplets around $\omega=n\,\omega_{\rm S}$, where $\omega_{\rm S} =2\pi/\Delta t$ and $n$ denotes integers. The multiplets have the equal splitting of the orbital frequency, $\Omega$ ($\Omega/\omega_{\rm S}=5.69\times 10^{-5}$ in Fig.\,\ref{fig:2.5}). 

Strictly speaking, each multiplet has infinite peaks. However, the tiny amplitude peaks are practically undetectable, and each multiplet looks like a finite number of components; the first Nyquist alias looks like a triplet, the second looks like a quintuplet, and so on. The degree of the apparent multiplicity becomes higher with the increase of $n$, because the Bessel coefficients with $n\ne 0$ become important with the increase of $\omega$.

The amplitudes of the peaks in the individual multiplet are determined by $\omega_{\rm S}\tau$. Since $\tau \simeq 190\,{\rm s} \simeq 2.20\times 10^{-3}\,{\rm d}$, 
$\omega_{\rm S}\tau = 0.663$, $2\omega_{\rm S}\tau = 1.326$, $3\omega_{\rm S}\tau = 1.989$,  $4\omega_{\rm S}\tau = 2.652$, $5\omega_{\rm S}\tau = 3.315$, $6\omega_{\rm S}\tau = 3.978$,
$7\omega_{\rm S}\tau = 4.641$, and $8\omega_{\rm S}\tau = 5.304$.
The Bessel coefficients at these values are graphically seen in Fig.\,\ref{fig:2.4}.

\begin{table}
\caption{Parameters for the {\it Kepler} LC sampling.}
\begin{center}
\begin{tabular}{cc}
\hline\hline
parameters & values \\
\hline
$\tau$ & $2.20 \times 10^{-3}\,{\rm d}$ \\
$\Delta t$ & $2.10\times 10^{-2}\,{\rm d}$ \\
$\omega_{\rm S}$ & $3.02 \times 10^2\,{\rm rad}\,{\rm d}^{-1}$ \\
$\Omega$ & $1.72\times 10^{-2}\,{\rm rad}\,{\rm d}^{-1}$ \\
\hline
\end{tabular}
\end{center}
\label{tab:2.3}
\end{table}%

\subsection{Multiplets of the $\mbox{\boldmath$Kepler$}$ LC data}
\label{sec:2.4}
It should be stressed here again that only the peak at $\omega=\omega_0$ with $n=0$ is a single peak. The aliases with frequencies $\pm \omega_{\rm S} \pm \omega_0$ look like a triplet, while those with $2\omega_{\rm S} \pm \omega_0$ are a quintuplet. The aliases associated with $3\omega_{\rm S}$, $4\omega_{\rm S}$, $5\omega_{\rm S}$, and $6\omega_{\rm S}$ look like a septuplet, a nonuplet, a undecuplet, and a tredecuplet, respectively.  


\section{Examples with simulated data}

\subsection{Importance of a large observational time span}

\begin{figure}
\begin{center}
\includegraphics[width=0.43\textwidth]{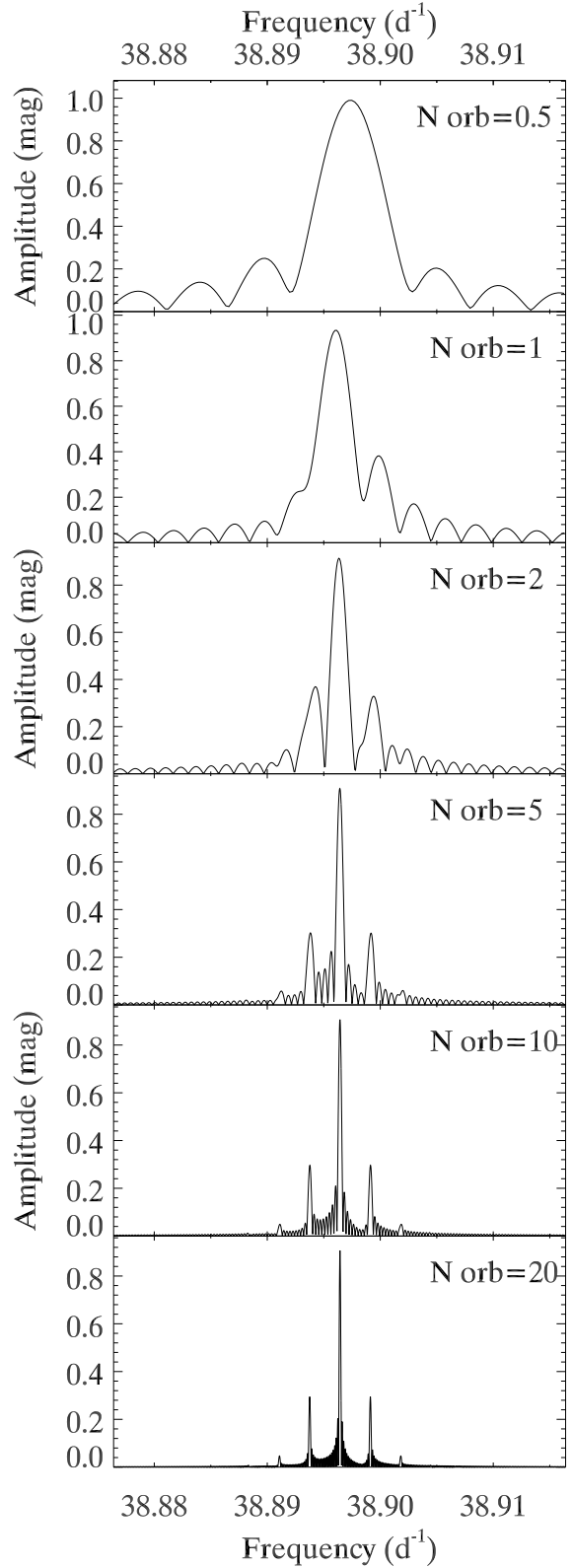}
\caption{Simulated data showing the frequency resolution achieved with different numbers of orbits of the satellite. For short observational time spans, the window pattern of each member of the multiplet distorts the shape of its neighbours. Calculated for Kepler field co-ordinates, at the first Nyquist alias of a 10\,d$^{-1}$ pulsation frequency.}
\label{fig:changing-norb}
\end{center}
\end{figure}

With the real \kepler\ data we are limited in our observational time span by the length of the mission thus far, assuming that the star we wish to study has been observed continuously since the mission began -- 3 complete orbits of the satellite have elapsed since then at the time of writing. However, with simulated data we are not so restricted. In Fig.\,\ref{fig:changing-norb} we show the importance of a large observational time span on resolving the multiplets into which Nyquist aliases are split. Although these simulated data are without noise, it is clear that when one orbit (four mission quarters) of data is available, aliases are no longer single peaks, but are split into sets of partially-resolved peaks and can thus be distinguished from real peaks (which are neither split nor similarly distorted). With two or more orbits, an obvious multiplet emerges, but the full-width-half-maximum of each peak is still large compared to the separation of the multiplet peaks, hence the window pattern of each peak can interfere with its neighbours. Increasing the number of orbits further sharpens the peaks and resolves the pattern for comparison with the analytical patterns we derived, shown in Fig.\,\ref{fig:2.5}.

\subsection{Effect of varying co-ordinates}

\begin{figure}
\begin{center}
\includegraphics[width=0.45\textwidth]{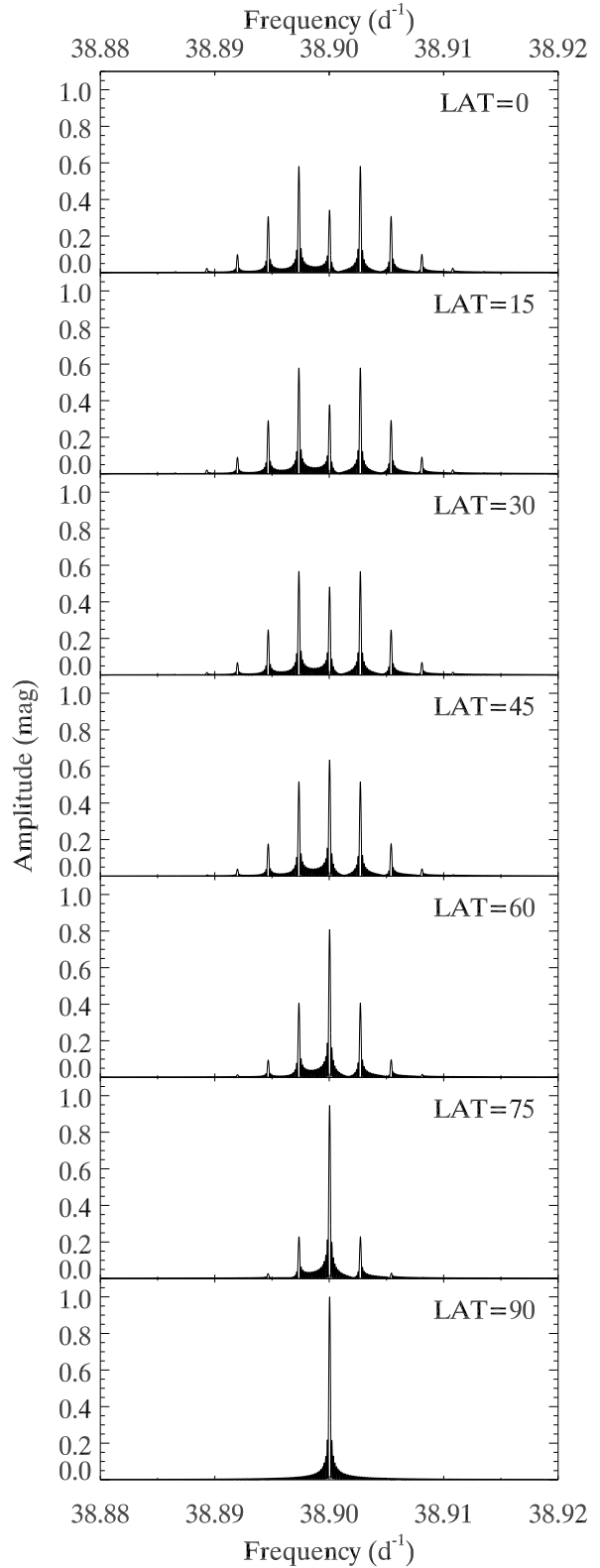}
\caption{Variation in multiplet shape for changing ecliptic latitude. Note that, because the ecliptic is inclined to the equator, varying either right ascension or declination will change the ecliptic latitude. At an ecliptic latitude of $90^\circ$, a single peak is seen because there is no longer any difference in light arrival time. Calculated for 20 orbits, at the first Nyquist alias of a 10\,d$^{-1}$ pulsation frequency.}
\label{fig:ecliptic-latitude}
\end{center}
\end{figure}

It is the barycentric time corrections that cause the splitting of Nyquist aliases. For a satellite that orbits around the Solar System barycentre in the ecliptic plane, these time corrections are determined by the ecliptic latitude of the target star, because this determines the difference in light arrival time at the satellite and the barycentre. We show how the multiplet shape changes with varying ecliptic latitude in Fig.\,\ref{fig:ecliptic-latitude}. In the more familiar celestial co-ordinates, both right ascension and declination affect the ecliptic latitude, so we expect the arrival times and thus the multiplet shape to change when either co-ordinate is varied. Since neither of these coordinates changes greatly over the \kepler\ field, the multiplet shape is almost identical, too.

\section{Super-Nyquist asteroseismology}

\subsection{Demonstration with real data}

We now compare our theory with real \kepler\ data by examining the case of a high-amplitude $\delta$\,Sct (HADS) star. These stars are the main-sequence counterparts to the classical Cepheids, having large amplitudes and normally few pulsation modes -- often the fundamental radial mode alone or with the first radial overtone. Sometimes harmonics of the frequencies of these modes are seen, as with the case we present here: that of KIC\,6861400. We used the Q1-9 LC data, which have been processed with the PDC-MAP pipeline \citep{stumpeetal2012, smithetal2012}. The star features three independent mode frequencies and many harmonics of the two mode frequencies with the highest amplitudes. We removed (`pre-whitened') all the harmonics from the data, and kept only the highest-amplitude mode of the three, leaving a single, high-amplitude peak to make our example. No data points were deleted.

\begin{figure*}
\begin{center}
\includegraphics[width=0.99\textwidth]{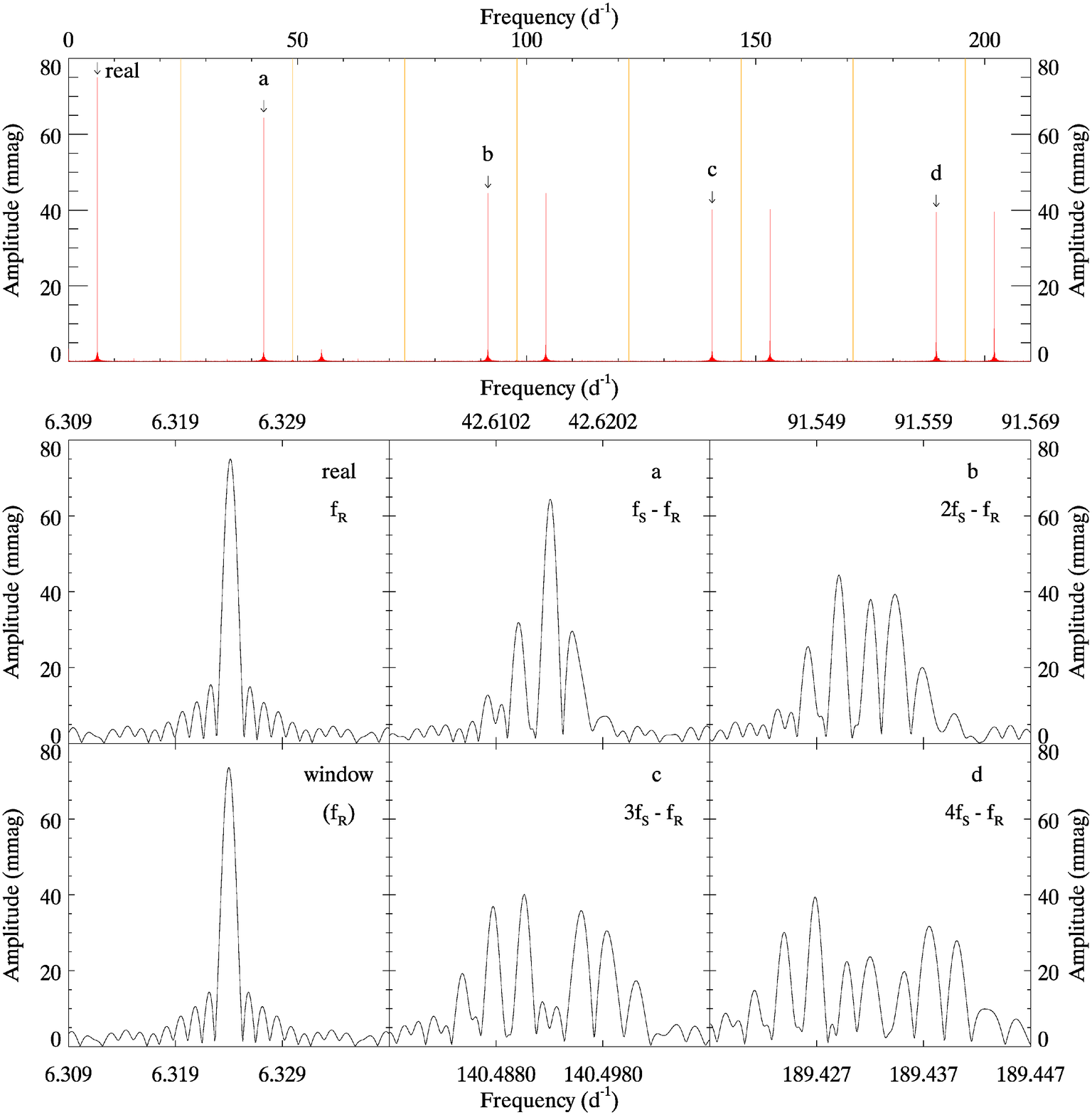}
\caption{The upper panel shows the frequency distribution of Nyquist aliases. In the lower panels, which are all to the same scale in frequency and amplitude, the left column shows the real peak, as labelled in the upper panel, and its window function. Panels a-d show aliases (also labelled in the upper panel), and their relationship to the \textit{sampling} frequency and the real frequency. The pattern of the number of peaks, $p$, as a function of the coefficient of the Nyquist frequency straddled, $n$, is apparent: $p = 2n + 1$. As such, there are three peaks at $f = f_{\rm S} + f_{\rm R}$, three at $f_{\rm S} - f_{\rm R}$, five each at $2f_{\rm S} \pm f_{\rm R}$ and so on, as shown in Fig.\,\ref{fig:2.5}. The example shown is KIC\,6861400, using Q1-9 PDC-MAP data.}
\label{fig:single-hads}
\end{center}
\end{figure*}

As expected, there is one alias in each frequency range $[nf_{\rm Ny}, (n+1)f_{\rm Ny}]$ (upper panel, Fig.\,\ref{fig:single-hads}), for integer $n$. Even values of $n$, counting from $n = 0$ at 0\,d$^{-1}$, are multiples of the sampling frequency. In Fig.\,\ref{fig:single-hads}, two peaks straddle each of these, and are identical in the Fourier transform. For this reason, only the peak to the left of each multiple of the sampling frequency is presented in the lower panels for a closer look. The multiplet structure is clearly present in all but the real peak, which is represented by a unique singlet peak. The spacing in the multiplets is approximately equal to the orbital frequency of the satellite, but due to the small number of orbits covered (2.25), the frequency resolution is insufficient to separate fully the window patterns of the members of the multiplets. The window pattern of each peak of the multiplet thus interferes with neighbouring multiplet members, and distorts the shape somewhat, as we showed in Fig.\,\ref{fig:changing-norb}.

As the frequency increases when we study consecutive aliases of higher $n$, the argument to the Bessel functions is also higher and the higher-order (higher $k$) terms become more significant. As such, the number of peaks seen increases. The amplitudes of the more distant members of the multiplet increase -- we begin to see the expected quintuplets, septuplets and so on (cf. \S\,\ref{sec:2.4}), as these peaks' amplitudes climb above the noise level. The amplitude ratios of the peaks with respect to each other also change. In panel a, the central peak is dominant, but it is absent in panel c, where only the window pattern of the first order sidelobes can be seen in its place. The amplitudes of the Bessel orders that determine these peaks' amplitudes were shown as a function of frequency in Fig.\,\ref{fig:2.5}.

The frequency dependence of the shape of the multiplet is on the sampling frequency straddled, only. The multiplet generated at the first Nyquist alias of a pulsation frequency at 20\,d$^{-1}$ looks exactly the same that of a 10-d$^{-1}$ frequency, and so on. Only the \textit{frequency} of the alias changes, not the multiplet shape.

\subsection{Distinguishing real peaks from Nyquist aliases}

\begin{figure*}
\begin{center}
\includegraphics[width=0.99\textwidth]{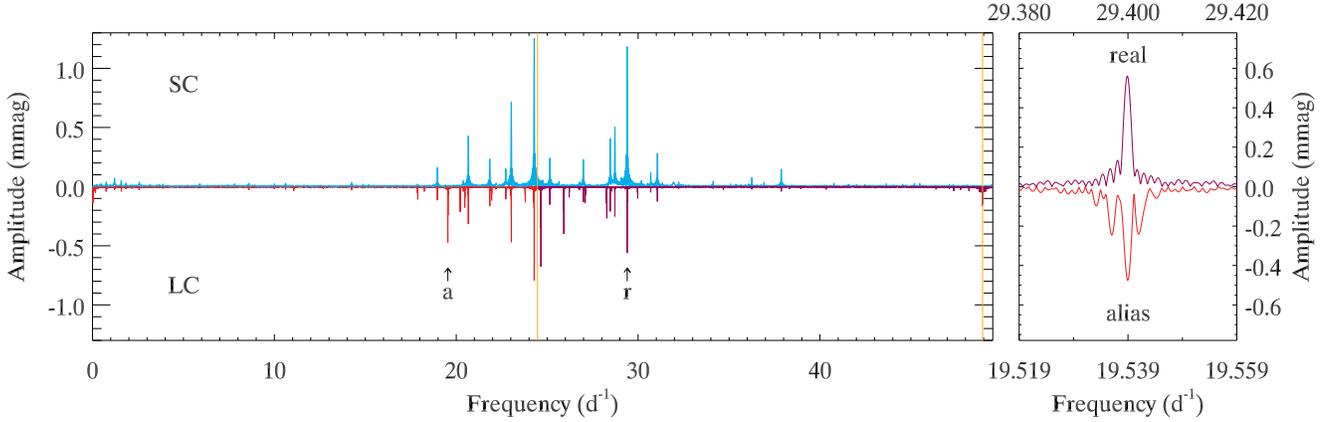}
\caption{In the left panel, the SC (top) and LC (bottom) data are displayed. LC Nyquist frequency and sampling frequency are represented with vertical orange lines. The LC data lying above their Nyquist frequency are coloured magenta; those below are red. There are Nyquist aliases in the LC data where real peaks have been aliased across the Nyquist frequency. By comparison with the SC data, we can see these aliases have no real partner in the SC data. An example of this is the alias peak labelled `a', corresponding to the real peak labelled `r' -- there is no peak in the SC data at the same frequency as peak `a'. We zoom in on `a' and `r' from the LC data in the right panel. The real peak takes on the typical sinc-function shape as expected in the Fourier transform. The effect of periodically-modulated sampling, however, is that the aliased peak is split into a multiplet, provided the time span of the data is at least one complete \textit{Kepler} orbit. The example given is the $\delta$\,Sct star KIC\,8590553. SC data cover Q4.3 only, but LC data are from Q0--9.}
\label{fig:realoralias}
\end{center}
\end{figure*}

The difficulty in observing $\delta$\,Sct stars with \kepler\ is that the typical pulsation frequency range of $\delta$\,Sct stars spans both sides of the LC Nyquist frequency. Those pulsation frequencies above the LC Nyquist frequency are aliased to lower frequencies in the Fourier transform, meaning that they overlap with the real peaks. Distinguishing these aliases from the real peaks is impossible with data that are truly equally spaced in time (cf. \S\,\ref{sec:2.1.2}), even if gaps exist in the data \citep{koen2010}. It has been argued that SC data are required to study the $\delta$\,Sct stars so that this problem of overlapping real and aliased frequencies is removed, because the SC Nyquist frequency is so much higher than the LC one. With a periodically-modulated sampling, however, one can use the shape of a peak to determine whether it is real or aliased. Fig.\,\ref{fig:realoralias} shows how a real peak exists as a single peak in the Fourier transform, but a Nyquist alias will be a multiplet (resolved or not) split by the satellite's orbital frequency.

Use of this technique to distinguish Nyquist aliases opens up asteroseismic possibilities for many {\it Kepler} targets that have been observed in LC only. With data sets that have time spans greater than one {\it Kepler} orbital period there is no Nyquist ambiguity in selecting the true pulsation frequencies for all types of pulsating stars. While there are still benefits of SC over LC data \citep{murphy2012}, the SC data are not needed to resolve Nyquist aliases.

\subsection{Beyond the Nyquist frequency}

The rapidly-oscillating Ap stars are a population of chemically peculiar A-type stars that pulsate with frequencies much higher than $\delta$\,Sct stars -- frequencies between about 70 and 250\,d$^{-1}$. As such, when viewed with LC data, these pulsations are many multiples of the Nyquist frequency away from falling into the range [0, $f_{\rm Ny}$]. It is still possible, however, to identify the true frequency ($f_{\rm R}=\omega_0/2\pi$) of the pulsations with LC data, even though the sampling period is much longer than the pulsation period, if the Fourier transform is calculated over a frequency range that includes $f_{\rm R}$.

\begin{figure*}
\begin{center}
\includegraphics[width=0.97\textwidth]{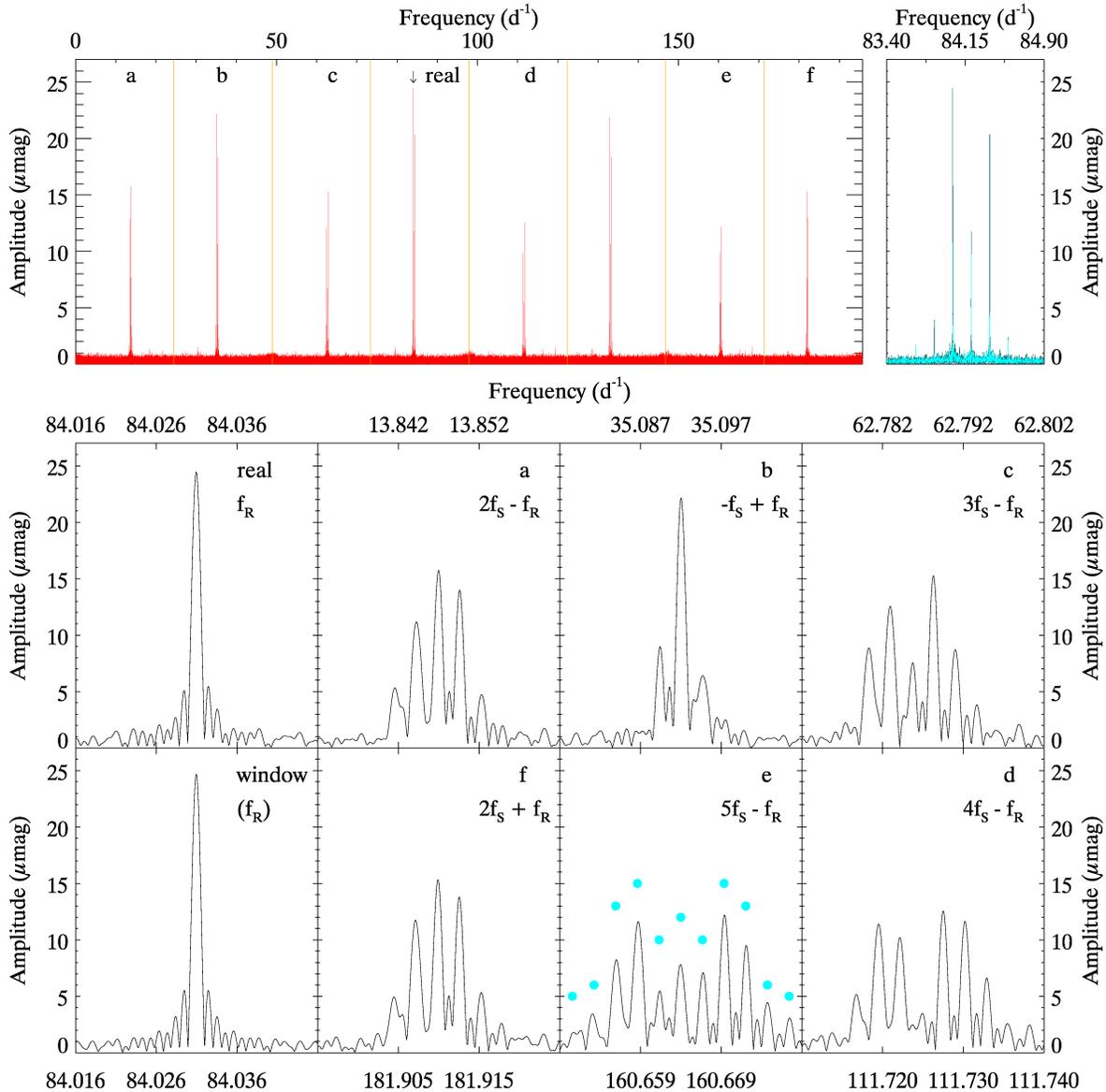}
\caption{In this example of the roAp star KIC\,10195926, a real multiplet of peaks appears around 84\,d$^{-1}$, and is shown zoomed in the top-right panel. The many aliases thereof are visible along with the real multiplet in the top left. The Nyquist frequency and its integer multiples are indicated as vertical orange lines. In the lower panels, we demonstrate the effect of periodically-modulated sampling on the shape of the highest-amplitude peak in each dipole-mode multiplet, all to the same scale. The real peak (far-left, top) has the typical sinc-function shape, even in these LC data and well beyond the Nyquist frequency. It can be compared to the window function in the far-left, bottom panel. The remaining lower panels, $a$-$f$, correspond to labels in the upper panel, and are in frequency order, clockwise. The data are Q1--11 LC data. See the text for further explanation.}
\label{fig:roAp}
\end{center}
\end{figure*}

We present an example in Fig.\,\ref{fig:roAp} using the star KIC\,10195926, whose roAp pulsations have periods among the longest known and whose frequency spectrum features an oblique dipole mode at $\sim84$\,d$^{-1}$ that is rotationally split into a septuplet \citep{kurtzetal2011}. We have run a high-pass filter on the data to prewhiten the low frequency content. Thus the Q1-11 LC data presented contain only the high frequency pulsation. With this figure we present a few final examples of the application of the technique to LC data.

Firstly, it is clear that despite being substantially higher in frequency than the Nyquist frequency, the real peak is recoverable and frequency analysis can be conducted without SC data. Secondly, peaks with the same relationship to the sampling frequency, i.e. the pairs matching $nf_{\rm S} \pm f_{\rm R}$, are split into multiplets of the same shape, e.g. the multiplets corresponding to $2f_{\rm S} - f_{\rm R}$ and $2f_{\rm S} + f_{\rm R}$ have the same shape, as is seen by comparing the lower panels $a$ and $f$. Thirdly, even though the peaks have low amplitudes (being only about an order of magnitude higher in amplitude than the noise), it is a testament to the quality of \kepler\ data and confirmation of the validity of our theory that we can see so many orders of each multiplet. Specifically, panel $e$ of Fig.\,\ref{fig:roAp} corresponds to the fifth Nyquist alias. We thus expect $5n+1 = 11$ equally-spaced peaks to be visible. We plot blue circles at intervals of (1/372.5)\,d$^{-1}$, i.e. the orbital frequency, from the central component of the multiplet. Each multiplet member can be seen to lie at its predicted position, beneath a blue circle. There are 11 blue circles, and all but the left-most one corresponds to a clear multiplet member. This left-most member is of low amplitude, barely above the noise level and is therefore not resolved from another noise or window-pattern peak. Its counterpart, under the right-most blue circle, is identifiable. If we had a longer observational time span the peaks would sharpen and cross-talk between their window patterns would decrease, likely leading to the final unresolved member becoming more distinguished.

\subsection{Distinction from other modulation}

Periodic amplitude and frequency modulation of a stellar pulsation signal both generate multiplets in the frequency domain. The critical factor that distinguishes the Nyquist aliases we describe here from frequency modulation multiplets in binary stars as described by \citet{shibahashi&kurtz2012}, or amplitude modulation multiplets that are seen in, for example, roAp stars and Blazhko RR\,Lyrae stars, is that the Nyquist alias multiplets are split by exactly the {\it Kepler} orbital frequency (1/372.5\,d), within the frequency resolution. It is exceedingly unlikely that other targets will have the same modulation frequencies, hence show the same splitting. In the unlikely event that such a coincidental frequency were found in some star, then careful examination of the sidelobe amplitudes and phases would clearly distinguish the cases. Section\,2 provides all the necessary information, should amplitude and phase need to be examined in this rare case. 

\section{Application to the Kepler SC data}
\label{sec:SC}
In the case of \kepler\ SC data, the sampling time interval is much shorter (at 58.9\,s) than in LC mode. Hence in SC we have:  $\omega_{\rm S}=9.217\times 10^3\,{\rm rad}\,{\rm d}^{-1}$, and then $\omega_{\rm S} \tau = 20.3\,{\rm rad}$ and $\Omega/\omega_{\rm S}=1.866\times 10^{-6}$. The SC Nyquist frequency is 734.07\,d$^{-1}$ (=8496.18\,$\mu$Hz).

Since the argument of the Bessel coefficients, $\omega_{\rm S}\tau$, is 30 times larger in the case of {\kepler} SC than that of LC, the multiplet becomes much higher order even in the case of the lowest Nyquist alias, and also the amplitude of each peak shrinks as the power is distributed into more peaks. In the case of the lowest Nyquist alias, at $\omega=\omega_{\rm S}$, the multiplicity is as high as 50, and the amplitude is reduced to about 20\:per\:cent of that of the true peak since the amplitude of the Bessel coefficients $J_k(\xi)$ asymptotically decreases as $\sim [2/(\pi\xi)]^{1/2}$. This means that, in the case of {\it Kepler} SC, each of the Nyquist aliases in the power spectrum look like a `forest' of peaks, while the true peak is a singlet  five times higher in amplitude than the `forest'. Hence the singlet true peak looks obviously and conspicuously different from the aliases, irrespective of whether the true frequency is higher or lower than the Nyquist frequency. The true peak should be more easily distinguished than in the LC case. 

\begin{figure}
\begin{center}
\includegraphics[width=\linewidth]{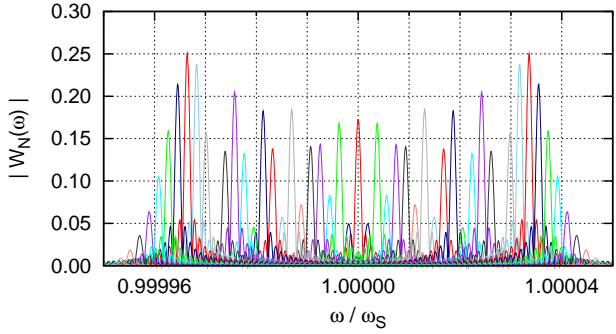}
\caption{The window spectrum for the lowest Nyquist alias at $\omega=\omega_{\rm S}$ of the {\it Kepler} SC data.}
\label{fig:sc-window}
\end{center}
\end{figure}

Fig.\,\ref{fig:sc-window} shows the window spectrum for the lowest Nyquist alias at $\omega=\omega_{\rm S}$ in the case of $N=10^5$. Most of the peaks therein are as short as 20\,per\,cent of the true peak, whose amplitude is normalised as unity, or much shorter. The multiplet would look like an unresolved, broad-band plateau if the resolution were lower.

\begin{figure}
\begin{center}
\includegraphics[width=\linewidth]{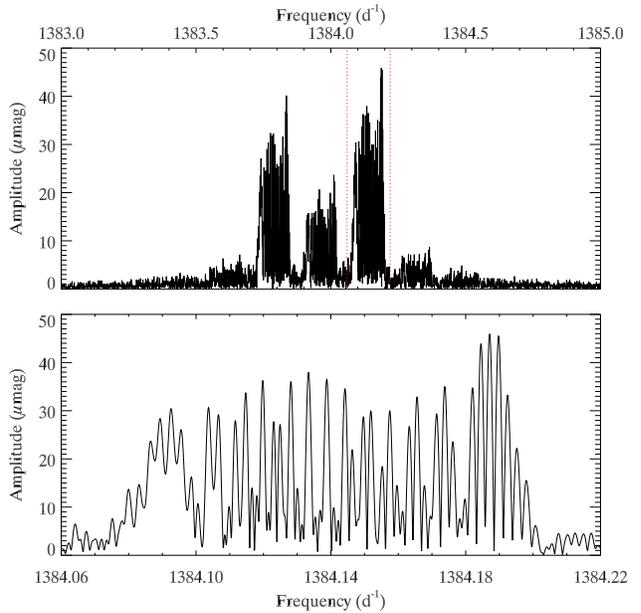}
\caption{The Nyquist alias of {\it Kepler} SC data of the roAp star KIC\,10195926. The frequency of this alias is $f_{\rm S}-f_{\rm R}$. The upper panel shows what the mode looks like as a whole.
The region delimited by the dashed red lines is shown in the lower panel, zooming in on the `forest'.}
\label{fig:roAp-SC}
\end{center}
\end{figure}

As a real example, we show in Fig.\,\ref{fig:roAp-SC} the alias peak in the SC data of our previous roAp star example. The upper panel demonstrates the appearance of the oblique dipole mode in the frequency range beyond the Nyquist frequency. One can see that the Nyquist aliases are at least six in number, where the seventh is at the noise level. Each peak looks like a thick pillar. That these are aliases is obvious because there is a forest where there would be a single peak if we were looking at the real pulsation frequency. The dashed red lines in the upper panel show the region that is plotted in the lower panel. 

As seen in this zoomed-in plot, the shape of the `forest' matches reasonably well with the expected window spectrum shown in Fig.\,\ref{fig:sc-window}, in particular, for the right half (on the left some peaks are unresolved in the real data) the relative amplitudes of each peak fit well with the expectation, as does the amplitude ratio of the alias to the true peak: the multiplet has amplitudes that range from around 30\,$\mu$mag for the unresolved part, to 45\,$\mu$mag for the right hand side, while the amplitude of the true peak in the Q6-12 SC data is 168\,$\mu$mag. The anticipated amplitude reduction to around 20\,per\,cent is confirmed. 

We provide one more example: that of aliasing of a pulsation frequency above the SC Nyquist frequency. We use the sub-dwarf B star KIC\,10139564 that was examined by \citet{baranetal2012}. In this star, non-linear combinations of mode frequencies lead to combination frequencies above the SC Nyquist frequency. It should be stressed that these frequencies are still real frequencies that describe the light variations of the star. \citeauthor{baranetal2012} correctly identified aliases of a few of these in their figure 20. The aliases identified were of the form $f_{\rm R}-f_{\rm S}$, and their Fourier calculations were based on 626\,000 data points covering 462.5\,d of SC data. Here, we extend the analysis to $10^6$ data points covering 739\,d.

\begin{figure*}
\begin{center}
\includegraphics[width=\linewidth]{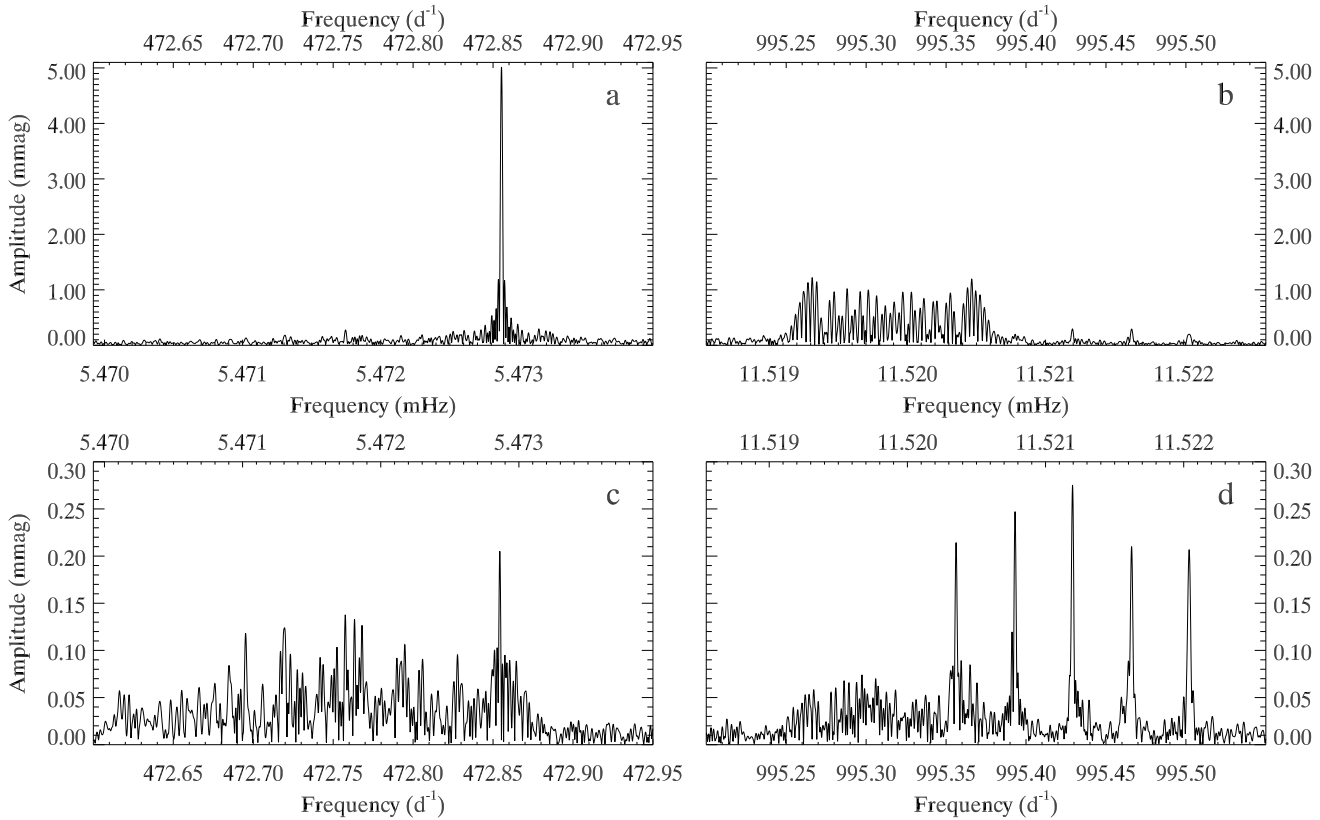}
\caption{Horizontally adjacent panels have the same amplitude scale; vertically adjacent panels cover the same frequency region. All panels are to the same scale in frequency. The tall peak of panel $a$ is the `$f_{18}$' of \citet{baranetal2012}, and is pre-whitened in the lower panels, \textit{only}. The Fourier transform is calculated on \kepler\ Q5--12 SC data, processed with pipeline version 8.0/8.1. Further explanation is provided in-text.}
\label{fig:sdB}
\end{center}
\end{figure*}

In addition to the combination frequencies at 995.43\,d$^{-1}$ reported by \citet{baranetal2012}, lies a Nyquist alias centred at 995.31\,d$^{-1}$. This frequency to which this Nyquist alias belongs was pre-whitened by those authors (identified there as $f_{18} = 472.855$\,d$^{-1}$) and so its Nyquist alias does not appear in their figure 20. We present the situation in Fig.\,\ref{fig:sdB}.  

Panel $a$ shows `$f_{18}$' of \citet{baranetal2012}. In lower panels only, $f_{18}$ has been pre-whitened. Its Nyquist alias, $f_{\rm S} - f_{18}$, is the broad forest in panel $b$, of substantially lower amplitude than the real peak, and in accordance with our theoretical expectations regarding both location in frequency and reduced amplitude. Also visible on the right of panel $b$ are some low-amplitude peaks that are combination frequencies in this sdB star. If $f_{18}$ is prewhitened, its alias multiplet disappears from panel $b$; what remains is visible in panel $d$ and includes those combination frequencies. Finally, those combination frequencies have Nyquist aliases of their own, but since those (real) combination frequencies are so low in intrinsic amplitude, their Nyquist aliases just contribute to the noise left in panel $c$ -- \citeauthor{baranetal2012} noticed and commented upon this increased noise, describing it as `low-amplitude complex signal'.

What we are seeing is two sets of frequencies ($f_{18}$ and the combination frequencies), that lie almost equidistant from and on opposite sides of the SC Nyquist frequency, such that the aliases of one set fall right next to the real frequency of the other set, and vice versa. Panel $b$ in particular highlights how easily these Nyquist aliases can be distinguished from real peaks, even (and indeed especially) in this very high frequency regime.


\section{Conclusions}

The barycentric time corrections applied to \kepler\ data break the regularity of the time interval between consecutive observations -- \kepler\ data are not equally spaced in time. For data spanning at least one orbital period, periodically-modulated sampling causes multiplets to be generated out of Nyquist aliases, whereas real peaks remain as singlets, irrespective of whether they lie above or below the Nyquist frequency. Multiplicity of the Nyquist aliases, along with relative amplitudes between multiplet members, depends on the number of multiples of the Nyquist frequency crossed, the amplitude of the sampling modulation, and the observational time span of the data.

We have theoretically derived the expected shapes of the alias multiplets and shown that the observed patterns are in agreement with the theory. We investigated the shape as a function of the number of multiples of the Nyquist frequency crossed, and obtained the following results: In real \kepler\ LC data, where noise is small but none the less present, the Nyquist aliases are split into multiplets whose members number $2n+1$, where $n$ is the number of multiples of the Nyquist frequency crossed. In the SC case, multiplicity is substantially higher and amplitudes are further reduced. Multiplet members are equally separated in frequency by the orbital frequency of the satellite in both cadences. The shape of the multiplet is independent of the actual pulsation frequency.

The resulting amplitude reduction upon aliasing precludes this method from being applicable to frequencies of very low inherent amplitudes, such as those of solar-like oscillators -- the amplitude reduction leaves no significant signal. The major consequence of our findings is that LC data may be used to study $\delta$\,Sct stars and roAp stars with real pulsation frequencies above the LC Nyquist frequency. This is also true using SC data for much higher frequency pulsators such as sdBV stars and pulsating white dwarf stars that have frequencies above the SC Nyquist frequency. The requirement is that continuous data sets are needed that have time spans greater than the orbital period of {\it Kepler}. This is the case for many hundreds of stars observed by {\it Kepler} in LC that previously were thought to have insurmountable ambiguities in their frequencies. 

\vspace{5mm}
\section*{Acknowledgements}

SJM acknowledges the financial support of the STFC. This work was carried out with partial support from a Royal Society UK-Japan International Joint program and also from a JSPS Japan-UK Joint Research project. We thank P.~Degroote and S.~Bloemen for their useful comments.

\bibliography{arxiv_sNa}

\end{document}